# Modified Bohmian quantum potential due to the second quantization of Schrödinger equation

Mohammad Jamali[1,a,b], Mehdi Golshani[a,b], Yousef Jamali[c]

[a]Department of Physics, Sharif University of Technology, Tehran, Iran
[b]School of Physics, Institute for Research in Fundamental Sciences (IPM), Tehran, Iran
[c]Department of Applied Mathematics, School of Mathematical Sciences, Tarbiat Modares University, Tehran, Iran

[1]Email address: byasinjamali@ipm.ir

# Modified Bohmian quantum potential due to the second quantization of Schrödinger equation

**Abstract**

A causal interpretation of the quantum world needs second quantization in order to cover phenomena like creation and annihilation of particles, and this leads to quantum field theory. The causal effects of the second quantization can be described through a new quantum potential. In this article we have dealt with the second quantization of Schrödinger and its effects on the path of a particle. This generalization leads to a modified Schrödinger which affects the particle through a modified quantum potential and a new term in the continuity equation. We have shown that these effects can provide a framework for the explanation of the creation and annihilation phenomena and other effects of quantum field theory on the development of a particle.

**Keywords**: modified Schrödinger equation, quantum potential, quantum field theory, second quantization

## 1. Introduction

During the last decade, there have been worthy efforts to extend Bohmian quantum mechanics to a quantum field theory [1-6]. Specifically, this has been done for the case of bosonic particles obeying Klein-Gordon equation [2, 3]: When we extend Bohmian mechanics to a quantum field theory, a new term appears in the Klein-Gordon equation [7, 8]. This new term is called quantum potential.

In the first quantization, a quantum potential appears that has all effects of the quantum level on the classical level and the path of particles. Bohm considers higher levels as well [9]. Higher levels can have effects on this quantum level through a new quantum potential. This higher level can be reached through a functional approach to quantum field theory, which can be called second quantization. Since the laboratory setups are arranged for detecting particles, we expect that all observable effects of the quantum field theory to be seen in this generalized quantum potential.

In this article, we started with the Schrödinger equation and quantized it to obtain the quantum potential at the level of the Schrödinger equation. This modified Schrödinger equation can generate a new quantum potential in the Hamilton-Jacobi equation, which has effects on the path of the particles. The new continuity equation also includes a new term, which is due to a second quantization and can explain phenomena like creation and annihilation of the particles, and the lack of conservation of probability at the level of parts, although the total probability is conserved.

## 2. Schrödinger picture of QFT and a modified Schrödinger equation in QM

To obtain the Schrödinger equation of the standard QM, we can take the following Lagrangian density (assuming that $m = \hbar = 1$):

$$\mathcal{L} = i[\psi\dot{\psi}^* - \psi^*\dot{\psi}] + 2\psi^* U\psi + \nabla\psi\nabla\psi^* \tag{SA.1}$$

Where "$U = U(x,t)$" is the classical potential. Using the action principle, the Euler- Lagrange equation is obtained:

$$\frac{\delta\mathcal{L}}{\delta\psi^*} - \frac{\partial}{\partial x^\mu}\frac{\delta\mathcal{L}}{\delta\psi^*_{,\mu}} = 0 \tag{SA.2}$$

$$\frac{\delta\mathcal{L}}{\delta\psi} - \frac{\partial}{\partial x^\mu}\frac{\delta\mathcal{L}}{\delta\psi_{,\mu}} = 0$$

Which leads to the Schrödinger equation:

$$-i\dot{\psi} + U\psi - \frac{\nabla^2\psi}{2} = 0 \tag{SA.3}$$

By definition. The corresponding (conjugate) momenta of $\psi^*$ and $\psi$ are, respectively, equal to:

$$\Pi = \frac{\delta\mathcal{L}}{\delta\dot{\psi}^*} = i\psi \tag{SA.4}$$

$$\Pi^* = \frac{\delta\mathcal{L}}{\delta\dot{\psi}} = -i\psi^* \tag{SA.5}$$

Now, we can obtain the Hamiltonian density:

$$\mathcal{H} = \Pi\dot{\psi}^* + \Pi^*\dot{\psi} - \mathcal{L} = -2\psi^* U\psi - \nabla\psi\nabla\psi^* \tag{SA.6}$$

$$H = \int d^3x\,\mathcal{H} = \int d^3x(-2\psi^* U\psi - \nabla\psi\nabla\psi^*) \tag{SA.7}$$

and calculate the Hamilton-Jacobi equation by using the identity of "$H + \dot{S} = 0$" and relations (4) and (5):

$$\dot{S} - \int d^3x(2\Pi^* U\Pi + \nabla\psi\nabla\psi^*) = 0 \tag{SA.8}$$

Now, using the identities:

$$\frac{\delta S}{\delta\psi^*} \equiv \Pi, \qquad \frac{\delta S}{\delta\psi} \equiv \Pi^* \tag{SA.9}$$

The Hamilton-Jacobi equation is obtained:

$$\dot{S} - \int d^3x\left(2\frac{\delta S}{\delta\psi}U\frac{\delta S}{\delta\psi^*} + \nabla\psi\nabla\psi^*\right) = 0 \tag{SA.10}$$

So far, everything has been classical. Now, according to the canonical method of second quantization [10], we can obtain the time evolution of the functional field. The canonical method includes the time evolution principle and the change of momenta to the field differential operators:

$$i\frac{\partial}{\partial t}\Psi = H\Psi \; ; \; \Pi \to \frac{i}{\hbar}\frac{\delta}{\delta\psi^*} \; , \Pi^* \to \frac{i}{\hbar}\frac{\delta}{\delta\psi} \tag{SA.11}$$

Now, by using equation (7) for the Hamiltonian, we have:

$$H = \int d^3x(-2\Pi^*U\Pi - \nabla\psi\nabla\psi^*) = \int d^3x\left(2\frac{\delta}{\delta\psi}U\frac{\delta}{\delta\psi^*} - \nabla\psi\nabla\psi^*\right) \tag{SA.12}$$

So, according to relation (11), the time evolution of functional field is equal to:

$$i\frac{\partial}{\partial t}\Psi = \left[\int d^3x\left(2\frac{\delta}{\delta\psi}U\frac{\delta}{\delta\psi^*} - \nabla\psi\nabla\psi^*\right)\right]\Psi \tag{SA.13}$$

Now, as in Bohmian approach, we can write the wave functional "$\Psi$" in the polar form in term of two real functional fields. Then, by separating the real and imaginary parts of equation ($SA.13$), we obtain the following equations:

$$\Psi(\psi, t) = \mathcal{R}(\psi, t)e^{i\mathcal{S}(\psi, t)} \tag{SA.14}$$

$$-\dot{\mathcal{S}} = \int d^3x\left(-\nabla\psi\nabla\psi^* - 2\frac{\delta\mathcal{S}}{\delta\psi}U\frac{\delta\mathcal{S}}{\delta\psi^*} + \frac{2}{\mathcal{R}}\frac{\delta}{\delta\psi}U\frac{\delta}{\delta\psi^*}\mathcal{R}\right) \tag{SA.15}$$

$$\frac{\dot{\mathcal{R}}}{\mathcal{R}} = \int d^3x\left[\frac{2U}{\mathcal{R}}\left(\frac{\delta\mathcal{S}}{\delta\psi}\frac{\delta\mathcal{R}}{\delta\psi^*} + \frac{\delta\mathcal{R}}{\delta\psi}\frac{\delta\mathcal{S}}{\delta\psi^*}\right) + 2\frac{\delta}{\delta\psi}U\frac{\delta}{\delta\psi^*}\mathcal{S}\right] \tag{SA.16}$$

Assuming the following identities:

$$\Pi \equiv \frac{\delta\mathcal{S}}{\delta\psi^*} \; , \qquad \Pi^* \equiv \frac{\delta\mathcal{S}}{\delta\psi} \tag{SA.17}$$

And inserting them in Eq. (SA.15), we get:

$$-\dot{\mathcal{S}} = \int d^3x\left(-\nabla\psi\nabla\psi^* - 2\Pi^*U\Pi + \frac{2}{\mathcal{R}}\frac{\delta}{\delta\psi}U\frac{\delta}{\delta\psi^*}\mathcal{R}\right) \tag{SA.18}$$

Now, by comparing this identity with the Hamilton-Jacobi equation of the classical field, Eq. (SA.8), a modified Hamilton-Jacobi equation is obtained, through the second quantization:

$$\dot{\mathcal{S}} = \int d^3x(2\Pi^*U\Pi + \nabla\psi\nabla\psi^* - 2\mathcal{Q}) \tag{SA.19}$$

$$\mathcal{Q} = \frac{\left[\frac{\delta}{\delta\psi}U\frac{\delta}{\delta\psi^*}\mathcal{R}(\psi,t)\right]}{\mathcal{R}} \tag{SA.20}$$

Then, by using the identity "$H + \dot{\mathcal{S}} = 0$", we get the modified Hamiltonian in the form:

$$H = \int d^3x(-2\Pi^*U\Pi - \nabla\psi\nabla\psi^* + 2\mathcal{Q}) \tag{SA.21}$$

$$\mathcal{H} = -2\Pi^*U\Pi - \nabla\psi\nabla\psi^* + \frac{2}{\mathcal{R}}\left[\frac{\delta}{\delta\psi}U\frac{\delta}{\delta\psi^*}\mathcal{R}(\psi,t)\right] \qquad (SA.22)$$

Therefore, by using the relation between Hamiltonian and Lagrangian, Eq. (SA.6), the new Lagrangian is equal to:

$$\mathcal{L} = \Pi\dot{\psi}^* + \Pi^*\dot{\psi} - \mathcal{H} = i[\psi\dot{\psi}^* - \psi^*\dot{\psi}] + 2\psi^*U\psi + \nabla\psi\nabla\psi^* - \frac{2}{\mathcal{R}}\left[\frac{\delta}{\delta\psi}U\frac{\delta}{\delta\psi^*}\mathcal{R}(\psi,t)\right] \qquad (SA.23)$$

By the variation of this new Lagrangian in terms of $\psi$ or $\psi^*$, and using the Euler- Lagrange equation, the modified Schrödinger equation is obtained at the QM level:

$$i\frac{\partial}{\partial t}\psi(x,t) = \left[-\frac{\nabla^2}{2} + U(x,t)\right]\psi(x,t) + \frac{\delta}{\delta\psi^*}\mathcal{Q}\Big|_{\psi(x,t)} \qquad (SA.24)$$

$$\mathcal{Q} = \frac{\left[\frac{\delta}{\delta\psi}U\frac{\delta}{\delta\psi^*}\mathcal{R}(\psi,t)\right]}{\mathcal{R}} \qquad (SA.25)$$

### 3. Modified Bohmian quantum potential due to second quantization

We have obtained a modified Schrödinger equation at the QM level, which has an extra term "$\frac{\delta}{\delta\psi^*}\mathcal{Q}$" relative to the normal Schrödinger equation. To obtain the second quantized effects on the dynamic and evolution of a particle, Eq. (SA.24) can be rewritten as:

$$i\hbar\frac{\partial}{\partial t}\psi(x,t) = \left[-\frac{\hbar^2}{2m}\nabla^2 + U(x,t) + \frac{1}{\psi}\frac{\delta}{\delta\psi^*}\mathcal{Q}\right]\psi(x,t) \qquad (SA.26)$$

As in the Bohmian approach [7], we write the wave function in the polar form:

$$\psi(x,t) = R(x,t)e^{\frac{i}{\hbar}S(x,t)} \qquad (SA.27)$$

where R and S are real functions. Then, using the following identity in the Eq. (SA.26):

$$\frac{1}{\psi}\frac{\delta}{\delta\psi^*} = \frac{1}{R}\frac{\delta}{\delta R} + \frac{i\hbar}{R^2}\frac{\delta}{\delta S} \qquad (SA.28)$$

We obtain the following two real functions, which are equivalent to the modified Schrödinger equation (Eq. (SA.26)):

$$-\frac{\partial}{\partial t}S = \frac{(\nabla S)^2}{2m} + U - \frac{\hbar^2}{2m}\frac{\nabla^2 R}{R} + \frac{1}{R}\frac{\delta}{\delta R}\mathcal{Q}\Big|_{R,S} \qquad (SA.29)$$

$$\frac{\partial}{\partial t}R^2 + \nabla\left(R^2\frac{\nabla S}{m}\right) = -2\frac{\delta}{\delta S}\mathcal{Q}\Big|_{R,S} \qquad (SA.30)$$

Assuming that particle's speed is equal to "$\frac{\nabla S}{m}$", the first equation ($SA.29$) has the same form as in the classical Hamilton-Jacobi equation, with two extra terms:

$$-\frac{\partial}{\partial t}S = \frac{(\nabla S)^2}{2m} + U + \mathbb{Q} \qquad (SA.31)$$

$$\mathbb{Q} = -\frac{\hbar^2}{2m}\frac{\nabla^2 R}{R} + \frac{1}{R}\frac{\delta}{\delta R}\mathcal{Q}\Big|_{R,S} \qquad (SA.32)$$

The first term is the same as the normal Bohmian quantum potential, but the second term is a new Bohmian potential which is due to the second quantization. In fact, potential "$\mathbb{Q}$" includes all effects of QFT level and QM level on the particle trajectory (particle dynamics level).

### 4. Continuity equation and evolutional effects

Considering Eq. (SA.30), we see that it is the same as the normal Bohmian continuity equation [7] with an extra term, which is due to QFT effects and second quantization:

$$\frac{\partial}{\partial t}R^2 + \nabla\left(R^2\frac{\nabla S}{m}\right) = -2\frac{\delta}{\delta S}\mathcal{Q} \qquad (SA.33)$$

In the standard QM, the term of "$R^2$" is interpreted as the probability of the particle detection, after observation, and in the statistical many particle Bohmian QM, it can be interpreted as a particle density distribution (or probability distribution of particle's location). As it is seen in Eq. (SA.33), the extra term "$2\frac{\delta}{\delta S}\mathcal{Q}$", apparently causes the none conservation of probability. However, if we repeat the above calculations for "$\psi^*$" field, another continuity equation is obtained, which is:

$$\frac{\partial}{\partial t}\bar{R}^2 + \nabla\left(\bar{R}^2\frac{\nabla S}{m}\right) = -2\frac{\delta}{\delta S}\bar{\mathcal{Q}} \qquad (SA.34)$$

where the $\bar{\mathcal{Q}}$ is equal to:

$$\bar{\mathcal{Q}} = \frac{\left[\frac{\delta}{\delta\psi^*}U\frac{\delta}{\delta\psi}\mathcal{R}(\psi,t)\right]}{\mathcal{R}} \qquad (SA.35)$$

If we consider the anti-commutator relation between $\Pi^*$ and $\Pi$, the result is:

$$\left[\frac{\delta}{\delta\psi^*},\frac{\delta}{\delta\psi}\right]_+ = 0 \qquad (SA.36)$$

$$\bar{\mathcal{Q}} = -\mathcal{Q} \qquad (SA.37)$$

Therefore, due to Eq. (SA.34) and Eq. (SA.37), the continuity equation for "$\psi^*$" field, is equal to:

$$\frac{\partial}{\partial t}\bar{R}^2 + \nabla \left(\bar{R}^2 \frac{\nabla S}{m}\right) = +2 \frac{\delta}{\delta S} Q \tag{SA.38}$$

Comparison with Eq. (SA.33) and Eq. (SA.38) can be interpreted as the survival probability that the whole system (particles and antiparticles) will preserve. This is because the dynamical operations at the QFT level are unitary.

In Bohmian QM, this extra term can be taken as a basis for the causal description of creation and annihilation phenomena, and in the Bohmian interpretation, this term affects the magnitude of the pilot wave (which is interpreted as active information [11-13]). But, we know that the quantum potential which is its result, is independent of wave magnitude. However, the effect of quantum potential is valid as far as the wave magnitude is not zero. When due to the Eq. (SA.33), "R" becomes zero, then there is no pilot wave or active information. So why must we consider a particle with no wave function? This situation can be interpreted as an annihilation effect. And the reverse process for "$\psi^*$" can be interpreted as a creation effect.

We can see this effect in the Eq. (SA.26), where the new extra term in the modified Schrödinger equation is complex. This term is same as an effective potential term (and effective non-unitary Hamiltonian) and we know that mathematically an imaginary potential can destroy the conservation of probability in the case of Schrödinger equation[14], and in many parts of physics, it can be considered as a dissipative effect. We propose that because of this extra term, which leads to a nonlinear modified Schrödinger equation, there may be a possibility for the solution of measurement problem in QM (due to its nonlinearity and non-unitary evolution). In the Bohmian interpretation of mind and matter relation, which was proposed in [9], the effect of higher level possibilities on the lower level dynamics, can be tied to the causal transformation of active information to inactive information in a Bohmian QM, and also, the reduction of the wave function in the standard QM. This is achieved through the change of regular Schrödinger equation to a nonlinear modified Schrödinger equation, thanks to QFT effects.

For bosonic particle (or real field), to preserve the conservation of probability, the extra term of "$\frac{\delta}{\delta S} Q$" must be zero, therefore "$Q$" must be invariant from phase transition, and it causes a kind of charge conservation due to Noether's theorem.

## 5. Conclusion

In Bohmian mechanics, the effects of the level of possibilities, described by the wave function, on particle trajectories are summarized in the quantum potential. By the quantization of the Schrödinger equation, we obtain a functional field, which shows the possibilities of the wave function. We have shown that this generalization to quantum field theory leads to an extra potential-like term in the Schrödinger equation. We call this new equation, the modified Schrödinger equation. (Eq. (SA.24), (SA.25)).

It was shown that this modified Schrödinger equation causes two effects on the particle evolution. One is via the modified quantum potential (Eq. (SA.32)) which has a new extra term with aspect to the normal Bohmian potential. It determines the effect of second quantization on the particle trajectory. Another is an extra term in the continuity equation which can provide a basis for a causal explanation of QFT level effects on the particle evolution, such as creation and annihilation phenomena. We propose that this dissipative extra term maybe considered as a solution for the measurement problem in the standard QM, because of its effect on the nonlinearity of modified Schrödinger equation, and also providing a base for producing a mechanism of transformation of active information to inactive one in the Bohmian interpretation and its relation to the mind effect on matter.